\documentclass[preprint, secnumarabic, amssymb, amsmath,
aip, cha, frontmatterverbose]{revtex4-1} 
\usepackage{enumerate}
\usepackage{natbib} 
\usepackage[lflt]{floatflt}
\usepackage[latin1]{inputenc} 
\usepackage[pdftex]{graphicx}
\usepackage{color} 
\usepackage{amsmath} 
\usepackage{siunitx}

\newcommand{\tm}[1]{\textmd{#1}}

\begin{document}

\title{From nanoparticles to nanocrystalline bulk:\\ Percolation effects in field assisted
  sintering of silicon nanoparticles}

\author{D.\ Schwesig}
\affiliation{Department of Physics and Center for Nanointegration
Duisburg-Essen (CeNIDE), University of Duisburg-Essen, Germany}
\author{G.\ Schierning}
\author{R.~Theissmann}
\author{N.~Stein}
\author{N.~Petermann}
\author{H.~Wiggers}
\author{R.~Schmechel}
\affiliation{Faculty of Engineering and Center for Nanointegration
Duisburg-Essen (CeNIDE), University of Duisburg-Essen, Germany}
\author{D.\ E.\ Wolf}
\affiliation{Department of Physics and Center for Nanointegration
Duisburg-Essen (CeNIDE), University of Duisburg-Essen, Germany}

\begin{abstract}

Nanocrystalline bulk materials are desirable for many applications
as they combine mechanical strength and specific electronic
transport properties. Our bottom up approach starts with tailored
nanoparticles. Compaction and thermal treatment are crucial, but
usually the final stage sintering is accompanied by rapid grain
growth which spoils nanocristallinity. For electrically conducting
nanoparticles, field activated sintering techniques overcome this problem.
Small grain sizes have been maintained in spite of consolidation.
Nevertheless, the underlying principles, which are of high practical importance,
have not been fully elucidated yet.

In this combined experimental and theoretical work we show, how the
developing microstructure during sintering correlates to the
percolation paths of the current through the powder using highly doped
silicon nanoparticles as a model system.  It is possible to achieve a
nanocrystalline bulk material and a homogeneous microstructure.  For
this, not only the generation of current paths due to compaction,
but also the disintegration due to Joule heating is required.  The
observed density fluctuations on the micrometer scale are attributed
to the heat profile of the simulated powder networks.

\end{abstract}

\maketitle
\section{Introduction}

A growing demand for nanocrystalline materials is driven by
applications of energy materials e.g.\ for lithium batteries or fuel
cells \cite{Arico}, or for thermoelectricity \cite{ChenDresselhaus,
  Snyder} which make use of specific mechanical, electrical,
magnetic or optical properties of the nanomaterial. The
  bottom-up approach towards nanocrystalline bulk material starts from
  a nanoparticle powder and compacts it in a way that preserves the
  specific nanofeatures. Within the last years, this bottom-up
approach has become increasingly applicable for a growing number
of chemical compositions, as new reactor concepts and strategies can
supply tailored nanoparticles like zinc oxide, silicon, alumina, or
yttrium oxide in sufficient quantities.

The way towards a dense nanocrystalline sample is tricky\cite{Wei}.
A characteristic feature of nanopowders is their low tap density combined
with high stiffness, mainly originating from strong interparticle van-der-Waals forces.

The energy stored within the surface of the nanomaterial is a
driving force for coarsening as soon as temperature is applied. The
consequence is that the specific nanofeatures get lost.  Traditional
sintering methods like pressureless sintering and hot pressing
require the samples to be kept at high temperatures for a certain
time resulting in strong coarsening, so that the effort of
synthesizing nanosized starting particles is lost during the
sintering process. Sometimes, the over-all result is even inferior
to sintering micrometer-sized particles.

As a consequence, an old method has been rediscovered, compare review \cite{Munir}.
Field activated sintering techniques (FAST), also called
electric current sintering or spark plasma sintering, have
demonstrated their strength in sintering nanoparticles to dense bulk
under preservation of the nanogranular structure, while densities
near the one of the respective crystalline bulk material can be obtained \cite{Yoshimura}.
Field assisted sintering is a rather quick method, which makes it
also attractive for production processes. For both, hot pressing and
FAST, the compaction is done by a combination of pressing and
heating. The sample is located within a graphite crucible between
graphite dies. Those create reducing conditions. During the whole
process, a pressure is applied on the sample, usually in the order
between \SI{10}{\mega\pascal} and \SI{100}{\mega\pascal}, which, by itself, is
too small to compact the powder. For highly insulating
material, the basic difference between hot pressing and FAST is the
heating process. Field assisted sintering techniques use Joule
heating applied by driving a kA-current through the graphite
crucible, and ohmic heating increases the sample temperature from outside.
This creates a radial temperature field within
the sinter-body \cite{Vanmeensel, Dobedoe}. Still, the same main
mechanisms as observed for hot pressing, lattice diffusion and grain boundary
diffusion, dominate the sintering process\cite{Suganuma}.

If conducting bulk material is exposed to the electric current
heating, it is found that the temperature gradients within the
sample are much higher than in non-conducting material, shown by
temperature field simulations \cite{Vanmeensel}. If particles shall
be compacted using field assisted sintering, the intrinsic mechanism
of densification becomes a different one. Especially, if the process
is conducted in a way that the sample is electrically isolated from
the crucible and the path of the electric current is through the
sample preventing a short cut via the crucible.  Then percolation
effects inside the sample have to be considered. For a given
current through two particles in contact, most ohmic heat is
released, where the resistance is highest, which is generally
thought to be at the interface.  As a result, a partial melting at
the surface takes place as supposed by Groza et al.\ \cite{Groza1,
Groza2}.  After the current pulse, this partially molten zone will
have a lower resistance --- with back coupling on the current path.

The successful utilization of this process has been reported, but a
theoretical explanation how it works is still missing.  In a combined experimental
and theoretical approach, we have investigated the field-assisted
sintering of highly doped silicon nanoparticles. Nanocrystalline
silicon \cite{Bux, Schierning2010} and silicon based alloys
\cite{Joshi} are promising materials for thermoelectric
applications. Additionally, they are also a model system for a
conducting, covalently bound material. It has been shown by several
groups that nanocrystalline silicon can be densified applying field
assisted sintering techniques with very short sintering time in the
order of minutes \cite{Bux, Schierning2010}.

\section{Methods}

\subsection{Model of field assisted sintering }

Electrical currents through a nanopowder flow along a complex,
inhomogeneous and anisotropic network of paths, essentially determined
by the pore structure. Hence Joule heating is non-uniform and leads to
structural changes correlated with the structure itself in an intricate
way. These correlations between structural kinetics and local
structure are an important aspect of field assisted
sintering. They influence the outcome of the compaction process, as we
are going to show below.

To investigate such statistical correlation mechanisms of field assisted
sintering, a phenomenological network model may be used.
Densification, the current path through the powder, the obtained
temperature fields, and the structural evolution can be
calculated. In this approach material properties and processing
  enter only via a few thermodynamic and electrical
  parameters (e.g. specific heat, melting temperature, resistivity,
  sample cooling, pressure). Comparison with microstructures
obtained experimentally for different sintering parameters is crucial
for the validation of such a phenomenological model.

In this paper, cornerstones of such a network model will be
  proposed. For an exploratory study a two-dimensional model
  suffices. It should give qualitative insight, but does not allow
  quantitative comparison with experiments, which will require a
  three-dimensional version and further refinements of the model in
  the future.

The nanopowder is represented by a square grid, the nodes of
  which are partially filled with particles. On two opposite sides
  periodic boundary conditions are imposed. On the other two sides
  electrodes are attached to the powder. A certain electrical
  resistance is attributed to each bond connecting a particle to an
  electrode or to another particle. The top electrode can move
towards the bottom electrode to simulate the effect of applied
pressure.

Possible mechanisms of densification during sintering are
lattice diffusion, grain boundary diffusion, plastic deformation due
to dislocation activity, sliding of particles, or viscous flow
of completely or partially molten particles.  The fact that a
densification up to 97 \% of the maximally achievable density
is obtained within the short period of three minutes
\cite{Schierning2010} rules out the slow diffusion mechanisms
for field assisted sintering.  In a covalently bound material, the
mobility of dislocations and grain boundaries can be neglected, too.
Therefore, the most plausible model for the description of
silicon nanoparticle sintering should be based on viscous flow.

There are a couple of assumptions that may describe viscous flow
during silicon nanoparticle densification: i) It has been demonstrated
by differential scanning analysis (DSC) and in-situ transmission
electron microscopy (TEM) that silicon nanopowders comparable in size
and morphology partially melt and recrystallize upon heating at
temperatures below \SI{1000}{\celsius}.  It was observed, that
such a
partially molten powder could easily flow\cite{Schierning2008}.  ii) A
surface melting induced by the FAST process would result in a flow of
particles \cite{Groza1, Groza2}.  iii) It is a plausible assumption
that the silicon nanopowder is covered by a surface layer of native
oxide. It is assumed that the amorphous oxide becomes viscous at
elevated temperatures.

From the viewpoint of a model, it is of minor relevance, whether or
not the surface layer is an amorphous oxide shell or native silicon.
Both, molten silicon and softened silica would have a similar
effect. Even if the material transport would happen by
evaporation and redeposition \cite{Groza2}, the implementation into
the theoretical model would be the same, so that this case is
covered, too.

For all lines of argumentation, the integration into a simulation
model requires the same principal mechanism: Within the percolation
path of the current, there will be a constellation of neighboring
particles, where the ohmic heating will create the highest
temperature. Above a certain threshold, this hottest nanoparticle
will flow, due to complete melting, partial surface melting, or
viscous flow of a surface layer. The matrix surrounding this flowing
particle remains rigid, as the local temperature is lower there.

\begin{figure}[ht]
\centering
  \includegraphics[width=0.35\textwidth]{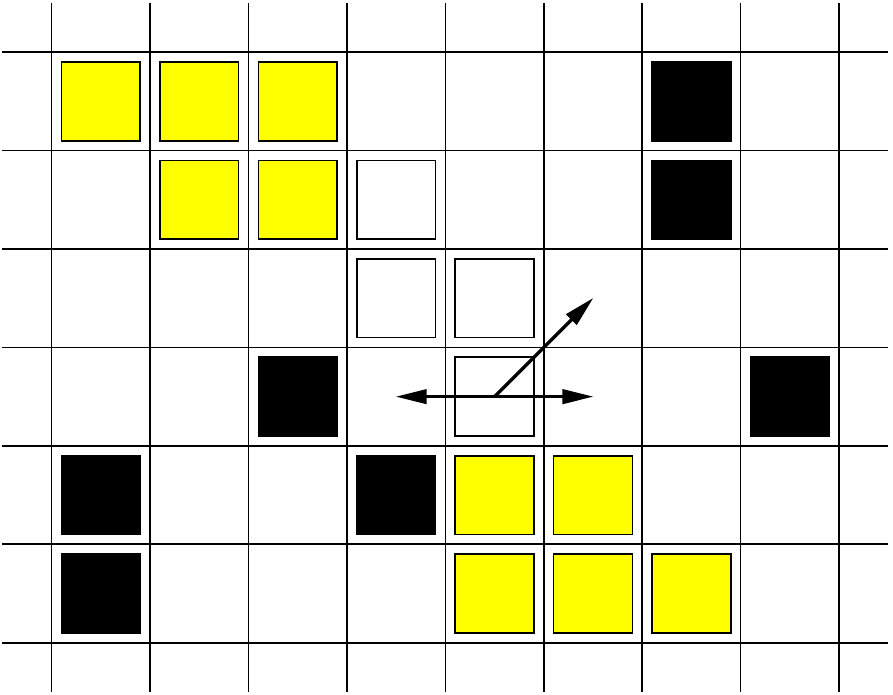}
  \caption{
  Possible movements of a flowing particle.
  The color coding gives a rough example of the temperature of the particle
  at the given time step.
  White squares represent molten nodes/particles, yellow color indicates nodes which have been heated up and black color represents cold nodes.
}
  \label{fig:MeltnMove}
\end{figure}

From in situ TEM investigations\cite{Schierning2008} we assume
that the flowing particle moves out of the current path. This
also generally reflects the behavior of liquid droplets connected to
solids (e.g.\ experienced when soldering). Within the real silicon
nanocomposite, a recrystallization at adjacent particles with lower
temperature will occur. If a native oxide shell is assumed to be the
viscous phase, a segregation of the oxide accompanies this
rearrangement. Within the simulation, we imagine that partially
molten particles move into a neighboring void as demonstrated in
figure \ref{fig:MeltnMove}. A stepwise rearrangement of the pores is
the consequence. If a complete horizontal path is emptied, the upper
electrode compacts the material by a movement downwards.

At this stage of model development we do not implement heat
  diffusion explicitly. Instead, we assume that temperature
  differences within the sample, if not renewed by Joule heating,
  would vanish after a relaxation time $\tau$. This limits the buildup
of temperature fluctuations in the sample. In the simulations we
take $\tau$ as discretization time.

\subsection{Simulation Parameters}

In this section, the simulation model will
be explained:

We simulate the powder on a two dimensional square grid of size
$L_{\tm{x}}\times L_{\tm{y}}=200\times 250$ with periodic boundary
condition in x-direction. The electrodes are the boundaries in
y-direction. Each node of the grid can be empty (representing pore
  volume) or occupied by one particle. Hence the fixed grid spacing
  can be identified with the particle diameter. Size differences between
  the particles are neglected. Particles sitting on nearest neighbor
  nodes (in x- or y-direction) are regarded as being in contact.

\begin{figure}[hb]
\centering
  \textbf{a}\includegraphics[width=0.24\textwidth]{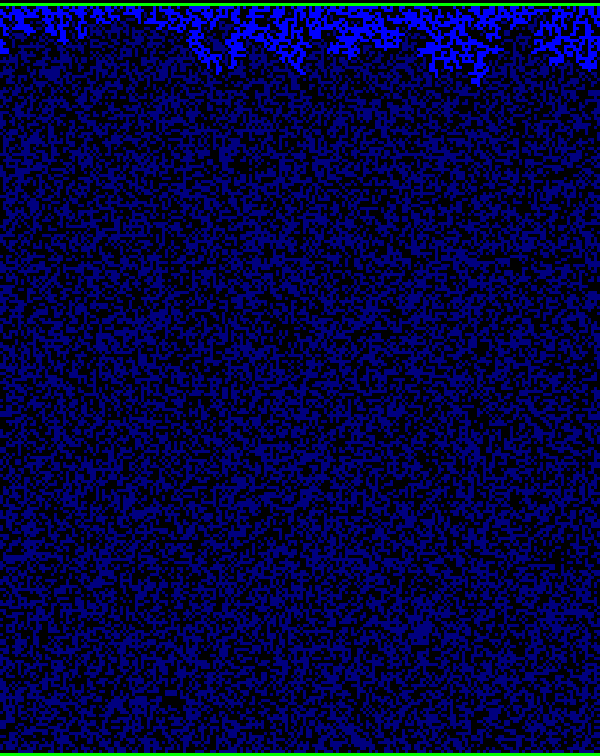}
  \textbf{b}\includegraphics[width=0.24\textwidth]{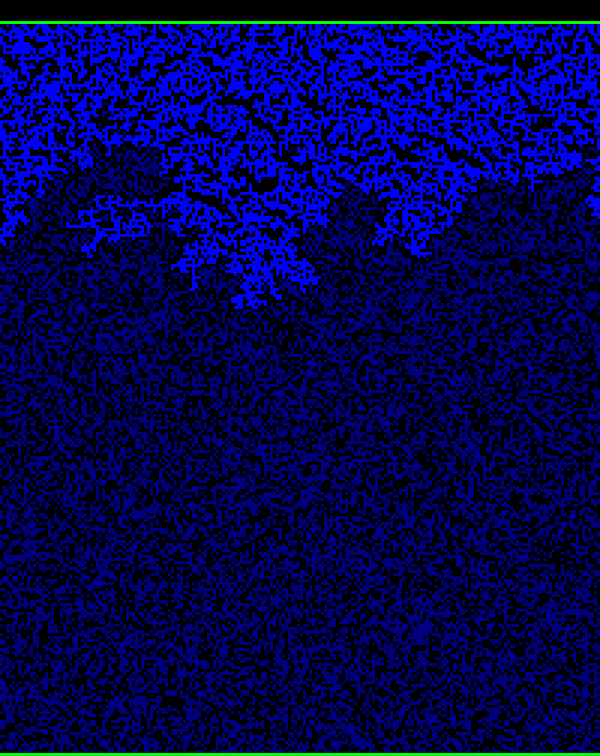}
  \textbf{c}\includegraphics[width=0.24\textwidth]{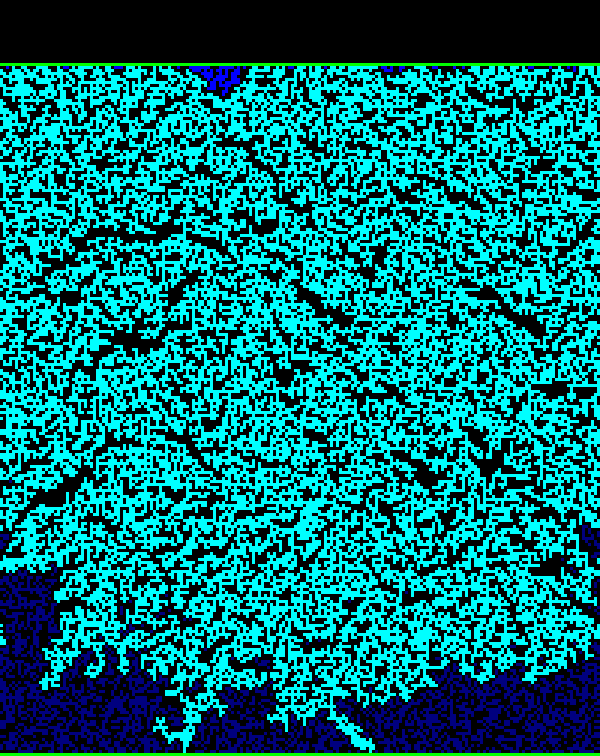}
  \caption{
  Two steps of the initial compaction until the powder percolates through the system.
  Empty nodes are colored black, the electrodes are green.
  The light blue nodes belong to the moving cluster with contact to the upper electrode.
  The dark blue nodes are occupied.
  The resulting percolating network is shown in picture c) in cyan.
  }
  \label{fig:Comp}
\end{figure}

The initial configuration, Fig.\ref{fig:Comp}c), is obtained by
  the following procedure: We start from a lattice gas, where half of
  the nodes are filled with particles at random positions,
  Fig.\ref{fig:Comp}a). The particles in contact to the upper
  electrode are shown in light blue. As long as there is no
  percolation path down to the lower electrode, the upper electrode
  can push the light blue cluster down by one grid spacing, without
  changing its structure. When this process is repeated, the cluster
  of particles in contact with the upper electrode grows
  (Fig.\ref{fig:Comp}b)), until the first percolating path forms,
which spans from the upper to the lower electrode
(Fig.\ref{fig:Comp}c)). It is assumed that the pressure applied to
  the upper electrode is so weak that a single percolating path is
  enough to resist further compaction. This is the initial
  configuration for the simulation of field assisted sintering. The
  initial inhomogeneity near the lower electrode is greatly reduced after a few
  simulation time steps as will be shown below.

The simulation proceeds as follows: The current through the
  system is calculated by a fast algorithm described in
\cite{ElmNode} for fixed voltage between the electrodes. The
  electrical resistance $R$ attributed to the contacts between
particles is taken as constant. The Joule heat $R I_{ij}^2
  \tau$ produced by the current $I_{ij}$ from particle $i$ to its
  neighbor particle $j$ during one time step $\tau$ is assumed to be
  delivered in equal parts to both particles, so that particle $i$
  receives the heat $\Delta Q_i = 0.5 \ \sum_j (R I_{ij}^2)\ \tau$.

We assume that the the sample is kept at a constant
  average temperature $T_{\tm{Sample}}$ and the heating compensates
  the energy loss to the environment at fixed $T_{\tm{envmnt}}$.
\begin{equation}
\frac{1}{N}  \sum_{i=1}^N \Delta Q_i = \left< \Delta Q \right>
\propto \tau \cdot \left( T_{\tm{sample}}-T_{\tm{envmnt}}\right)
\end{equation}
For $T_{\tm{sample}} >> T_{\tm{envmnt}}$ we get the approximation
\begin{equation}
  \left< \Delta Q \right> \propto T_{\tm{sample}}
\label{equ:avgQtoT}
\end{equation}
for the sample temperature.

Particle $i$ begins to melt, when the
heat $\Delta Q_i$ exceeds the threshold
\begin{equation}
 Q_{\tm{thres.}}  = C \cdot \left( T_{\tm{melt}}-T_{\tm{sample}}\right)
=C \cdot T_{\tm{sample}} \cdot \left( \frac{T_{\tm{melt}}}{T_{\tm{sample}}}-1 \right)
\end{equation}
needed to raise the local temperature up to the melting point of the
nanoparticles, $T_{\tm{melt}}$. The temperature dependence of heat
capacity $C$
close to melting has been neglected, and latent heat needs not be
taken into account, since the {\it onset} of melting is the
appropriate criterion for a particle displacement.

With equation (\ref{equ:avgQtoT}) we get the melting criterion
\begin{equation}
  \label{equ:Qmelt}
\frac{\Delta Q_i}{\left< \Delta Q \right>}  \ge m_{\tm{thres.}}
\propto  C \cdot \left(\frac{T_{\tm{Melt}}}{T_{\tm{Sample}}}-1
\right)  .
\end{equation}
This form is computationally particularly convenient, because the
ratio on the left hand side does not change, if the applied voltage
is rescaled. This has the advantage that $ \left< \Delta Q \right>$
does not need to be kept constant in eq.\ (\ref{equ:avgQtoT}). Any
change of $\left<\Delta Q \right>$ can be corrected by an appropriate rescaling of
the applied voltage without any effect on the simulation.

A particle with molten surface becomes mobile. It can be squeezed to
any free neighboring place as shown in figure
\ref{fig:MeltnMove}. This may interrupt the path of the electrical
current. If more than one particle have molten they move in random
order. Therefore it is possible that a disintegrated current path
instantly reappears. After the movement, the particles are reintegrated
into the rigid square grid.  The grid has a lower mean temperature
than the moving particles. Therefore, the reintegration physically
corresponds to a recrystallization \cite{Schierning2008}. We like
  to point out, that the model remains applicable, if instead of
  surface melting a temperature dependent viscosity of the surface
  layer leads to mobilization of hot particles and their subsequent
  arrest at a cooler neighboring place.

If the electrical contact between the electrodes was lost because of
particle movement, mechanical stability is lost as well. Then the
cluster of particles in contact with the upper
electrode is pushed down until percolation between
the electrodes reappears. Hence the sample becomes more compact.

Calculation of the currents, application of the
  melt criterion to move hot particles, and finally a compaction of
  the sample are repeated in every time step. The final configuration
  is reached, when the melt criterion is nowhere fulfilled any more or
  the sample became so dense that further compaction is impossible.
  Remarkably, the only model parameter is $m_{\tm{thres.}} \propto
  \left(\frac{T_{\tm{Melt}}}{T_{\tm{Sample}}}-1 \right)$. The
  importance of this key parameter for field assisted
  sintering has so far not been sufficiently appreciated.
  Its effect on the final microstructure
  will be investigated in the following. 

\begin{figure}[ht]
\centering
  \includegraphics[width=0.45\textwidth]{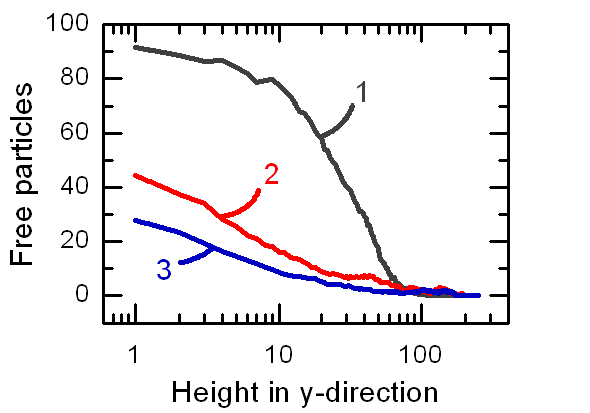}
  \caption{
    Free particles, e.g not connected to the spanning percolation
    cluster, versus the distance from the bottom electrode.
    The different curves are after different number of time steps
      of the sinter model algorithm for $m_{\tm{thres.}}=12$. Curve 1
    shows the initial situation. The inhomogeneity caused
    by the initial compaction is obvious. Curve 2 shows the
    situation after 50 time steps, curve 3 after 400 time
      steps. The initial increase of free particles towards the
      bottom electrode is strongly reduced already after 50 time steps.
  }
  \label{fig:Freepart}
\end{figure}

As pointed out above, the initial configurations for our
  simulations have a zone near the lower electrode that is poorly
  connected to the upper one (figure \ref{fig:Comp}c).  This is a
result of the method, by which the configurations were obtained, and
will not be found in a real powder.  We characterize this
inhomogeneity in figure \ref{fig:Freepart}, which shows the amount of
free particles which have
no contact to the upper electrode as function of the height
(y-direction) for different stages of the simulation.  As one can see,
the inhomogeneity is strongly reduced after a short period of about
50 time steps, but it does not vanish completely.  We assume that this has
only a minor effect on the path formation, because after the first 50
time steps the cluster is connected to the bottom electrode over the width
of the whole system, so that the path has enough freedom to
form.

  We use a length unit of \SI{20}{\nano\meter} for the grid to compare the simulation
pictures with the experimental TEM images of the sample.
This length corresponds to the average diameter of the nanoparticles
that were used in the experiment.

\subsection{Experimental Procedure}

Different batches of highly doped silicon nanoparticles with a
nominal doping concentration of \SI{5E-20}{\per\centi\meter^3}5 $\cdot$ 10$^{20}$ cm$^{-3}$ and a
mean starting diameter between \SI{20}{\nano\meter} and
\SI{40}{\nano\meter} were used for field assisted sintering
experiments, using a commercial system from FCT (Type HP D40, FCT
Systeme Rauenstein, Germany). System and process have been
analyzed e.g.\ in Ref.\ \cite{Vanmeensel}. Powders were precompacted
and usually they have a density of around 50 \% of the value of
crystalline silicon after this precompaction. Maximum temperatures
were set between \SI{860}{\celsius} and \SI{1160}{\celsius} at a
heating rate of \SI{100}{\kelvin\per\minute}, but it is known that
the measurement technique by pyrometer tends to underestimate the
actual temperature slightly \cite{Dobedoe}. Hold time was kept
constant at \SI{3}{\minute}. A pressure of \SI{35}{\mega\pascal} was
applied during the whole sintering process. Care was taken that the
sample was electrically isolated from the graphite crucible. This
was done by placing the sample within a graphite foil which was
coated by boron nitride from the outside. The current was thus lead
through the sample and not through the graphite crucible. The
process was temperature controlled using a pyrometer focussed on the
outer wall of the die wall surface. Current and voltage were set
according to the time-temperature program, current being in the
range of \SI{0.5}{\kilo\ampere} to \SI{1}{\kilo\ampere}, voltage in
the range of a few V.

Sintered nanocrystalline samples have about one third of the
phosphorus atoms electrically activated, resulting in a charge
carrier concentration around \SI{1E20}{\centi\meter^{-3}} and a specific
conductivity between \SI{100}{\siemens\per\centi\meter} and \SI{1000}{\siemens\per\centi\meter}.
At room temperature, typical values for the Seebeck coefficient are between \SI{100}{\micro\volt\per\kelvin} -- \SI{150}{\micro\volt\per\kelvin} and for the thermal conductivity between \SI{10}{\watt\per\meter\per\kelvin)} -- \SI{20}{\watt\per\meter\per\kelvin)}. The high specific conductivity allows
to conduct the experiment as described above.

For an investigation of the microstructure, samples were cut into
pieces and polished down to a thickness of around
\SI{40}{\micro\meter}. Final thinning to electron-transparency was
done using a precision ion polishing system by Gatan inc (PIPS 691).
Samples were characterized by transmission electron microscopy and
scanning electron microscopy (SEM), the latter using a special
adapter for TEM-samples. Microscopes used were a FEI Tecnai ST20 and
Jeol JSM 7500F, respectively. The geometric density was obtained by
the Archimedes principle by weighing the sample in ethanol and air.

\section{Results and Discussion}

\begin{figure}[ht]
\centering
  \includegraphics[width=1.0\textwidth]{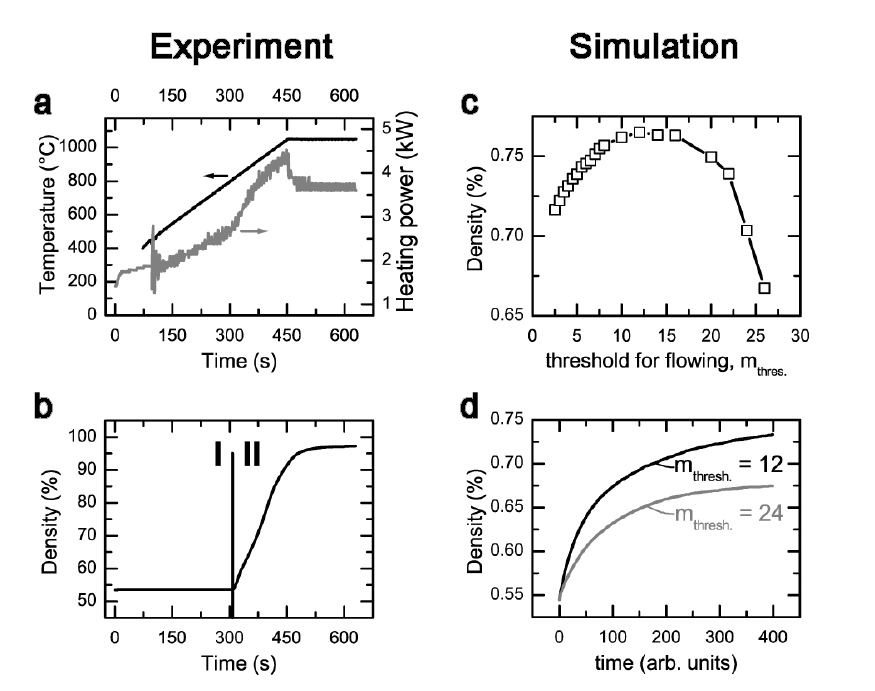}
  \caption{
  (a) Temperature $T$ and Joule heat $\dot Q$ delivered per unit
  time, vs. time. $T$ is measured by a pyrometer focused on the
  upper graphite die. $\dot Q$ is calculated from the applied current and voltage. The shown process is characterized by a constant
  heating rate of \SI{100}{\kelvin\per\minute} and a \SI{3}{\minute} hold time at \SI{1060}{\celsius}.
  (b) Density of the sample. The density is calculated from the movement of the
  upper graphite die and related to the end porosity derived from the geometrical
  density by Archimedes principal. Different phases of the process can be
  distinguished. See text for explanation.
  (c) Theoretical final densities received from
  simulations for different $m_{\tm{thres.}}$.
  (d) Densities development received from two different
  simulations with $m_{\tm{thres.}} = 12$, and
  $m_{\tm{thres.}} = 24$ for the densification (Phase II) of b).
}
  \label{fig:Process}
\end{figure}

In this section we compare the results of the simulation with
microstructural features found in the sintered experimental samples.
Figure \ref{fig:Process}a and \ref{fig:Process}b show an evaluation of the
sintering protocol during densification of a nanocrystalline sample. A
constant heating rate is followed by a hold time at \SI{1060}{\celsius}.
Two different phases can be distinguished:
\begin{enumerate}[{[}I{]}]
  \item {The first phase is characterized by a constant density of the
  sample. A rearrangement of particles is highly unlikely as it would directly
  affect the porosity. Any changes of the electrical
  properties of the sample in this phase could be caused by thermal and / or electrical
  breakdown of the oxide layer \cite{Xie1, Xie2} and local welding at the micro-contacts due
  to Joule heating as found for larger particles\cite{Falcon}.
  }
  \item {The second phase is characterized by a steep rise of density.
  Starting at the temperature of \SI{800}{\celsius}, the density increases
  from an initial value of 53 \% up to 97 \% caused by the sintering.
  This process slows down and the densification saturates eventually.
  The densification in this $2^{\tm{nd}}$ phase is captured by the simulation
  assuming particle reorganization due to local melting.
  }
\end{enumerate}
Around \SI{750}{\celsius} to \SI{800}{\celsius}, first melting
events were found for very similar batches of silicon nanoparticles
earlier \cite{Schierning2008}. Therefore, the temperature window
fits well to a fluid phase assisted sintering model assuming partly
or even completely molten silicon nanoparticles.

Figure \ref{fig:Process}d shows the simulation results for the
time evolution of the density for two different, constant sample
temperatures. They are represented  by the parameters
$m_{\tm{thres.}}=12$ and $m_{\tm{thres.}}=24$ chosen such that
the densification begins instantaneously. The density
increases and saturates
  to a value that depends on the sample temperature, in agreement with
phase II of the experiment.

Figure \ref{fig:Process}c) shows the simulated final density as
a function of $m_{\tm{thres.}}$. A low threshold represents a high
sample temperature that is close to the threshold temperature for
flowing. Accordingly higher thresholds are experimentally
realized by lower sintering temperatures. Obviously, for
high enough thresholds, no
  compaction takes place, in agreement with the low temperature
  behavior (phase I) of the experiment. Below a certain value of
$m_{\tm{thres.}}$ the density starts to increase. Surprisingly, the
simulation predicts that the compaction is maximal for a value of
$m_{\tm{thres.}}\approx 12$. For higher sample temperatures the
final density decreases again.  Experimentally it has so far only
been seen that with increasing sample temperature also the
compaction becomes stronger. Perhaps the high temperature behavior
cannot be seen in the experiment, because the melting temperature is
not uniform due to polydispersity. Another reason might be that
coordinated movement of particles was not considered in the
simulation, therefore a real material has more possibilities to
close pores.

\begin{figure}[h]
\centering
  \includegraphics[width=0.95\textwidth]{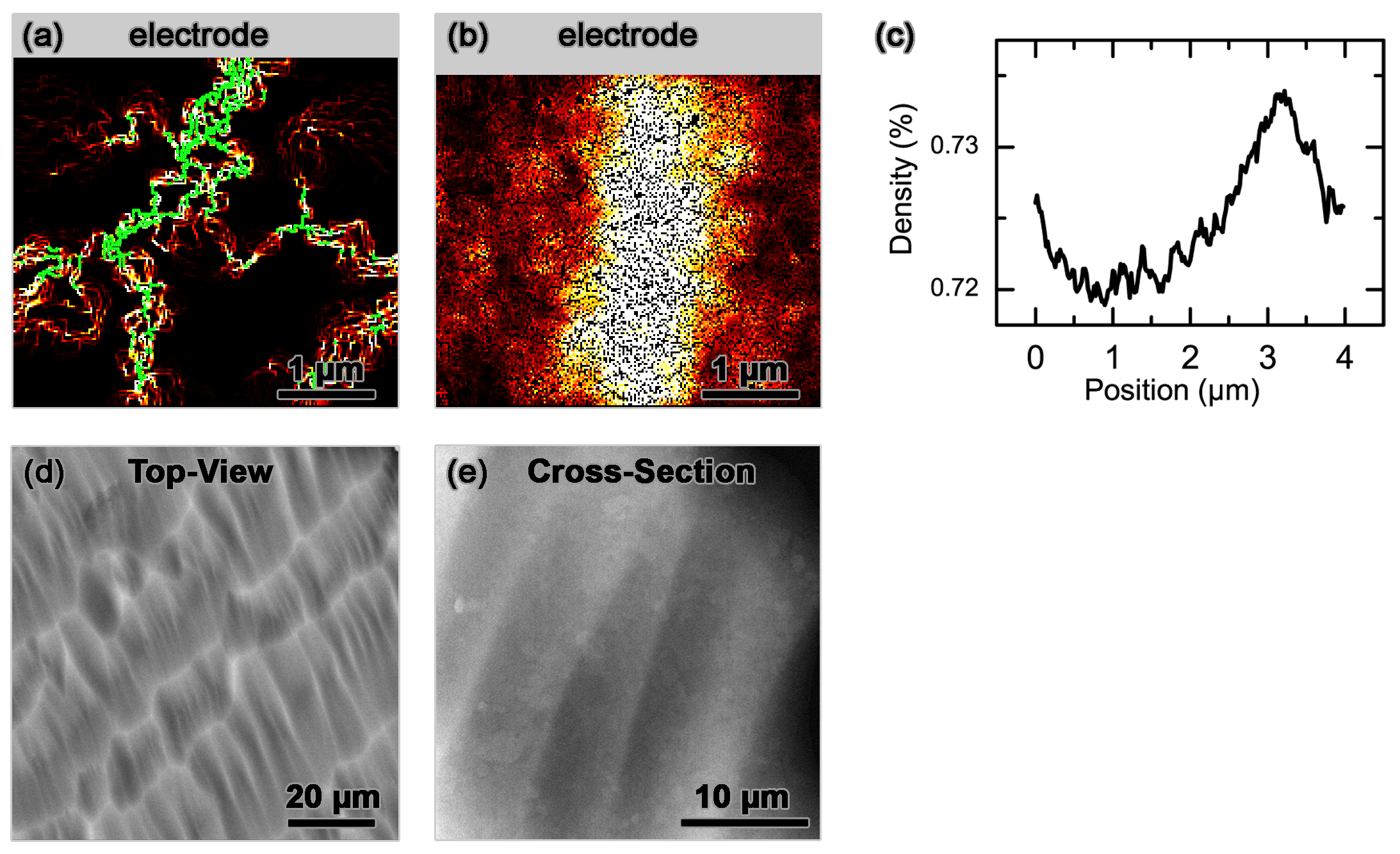}
  \caption{
    (a) Heat $\Delta Q_i$ generated  during the 100$^{th}$ time step
    of a simulated sintering process for $m_{\tm{thres.}}=5$. Particles
    plotted in green are above the threshold,
    have molten and will be moved as indicated in the text.
    (b) Accumulated heat at each particle after 400 time steps for for
    $m_{\tm{thres.}}=5$.
    (c) Density in a vertical slide through the system shown in b).
    To reduce the noise, the slide has a width of \SI{1.6}{\micro\meter}, which is in the same order as the path observable in b).
    (d) SEM image of a nanosilicon sample compacted by field assisted sintering
    (\SI{1060}{\celsius}, \SI{3}{\minute}). The same sample is shown as cross-section in e). Backscattered electrons
    are used for imaging to indicate fluctuations in density.
    Denser regions appear brighter.
  }
  \label{fig:Heat}
\end{figure}

The mechanism underlying the nonmonotonic temperature dependence
  of the final density is the following:
Let us consider cross sections of the simulated square grid parallel
to the electrodes. Most heat is deposited in the cross section,
which contains the least bonds, because this is the bottleneck of
the current. For large $m_{\tm{thres.}}$, selectively bonds in this
bottleneck melt until percolation between the electrodes is
disrupted and the sample comptifies. This continues until the
fluctuations in $\Delta Q_i$ become too weak so that the threshold
is no longer reached. Lowering $m_{\tm{thres.}}$ makes it easier for
compacted samples to still reach the threshold. However, this also
widens the molten region, which no longer is restricted to the
bottlenecks. This can be seen in Figure \ref{fig:Heat}a), where
basically a whole current path melts. The corresponding particle
displacement then amounts to a lateral diffusion of the percolating
path rather than its disruption. Hence compaction is hindered. This
explanation is supported by Figure \ref{fig:Heat}b), where the
accumulated heat is color coded to illustrate the diffusion of the
path.

While the development of the density seems to support our model,
the change of the electrical properties of the sample during the
ongoing FAST-process cannot be captured by our simulation. A
possible reason is that the moderate heating in phase I has not
been included in the simulation. Within this phase, the
percolation paths are not expected to disintegrate and fluctuate, as
disintegration would be accompanied by a rearrangement of the powder
network and thus a decrease of porosity which is not observed.
Instead, existing paths 'burn' into the sample. Contact resistances
within those paths may be optimized by breakthrough of the native
oxide layer and by local welding. This optimization creates initial
conditions for phase II which cannot be mimicked by constant contact
resistances between the particles as we have assumed within the
simulation. It can be concluded that the simulation so far
qualitatively captures the compaction of the field assisted
sintering of the highly doped silicon nanoparticles, but fails to
model the resistivity decay and increase, because of the constant
contact resistance.

In the following paragraphs, we will analyze the microstructures
obtained in field assisted samples.
We used the
contrast of backscattered electrons in SEM to make a fluctuation of the
sample's density visible (figure \ref{fig:Heat}c and d), assuming
that the atomic weight does not show a fluctuation\footnote{Note
that charging effects can also be excluded as origin of the observed
contrast because of the high doping level of the sample.}.
Obviously, there are paths within the nanosilicon sample which have
a higher density than the surrounding matrix, appearing at a
 micrometer length scale. In top view (figure \ref{fig:Heat} c) a
 honeycomb like structure is found.
In the crosssection linear paths are found, which
are very similar to the simulated heat pattern (figure \ref{fig:Heat} b).
Indeed, in the simulation a weak density increase correlated to this
heat pattern has been observed, Figure \ref{fig:Heat} c). Such patterns
are found regularly in nanosilicon samples
produced in our laboratory. We conclude that these regions of higher
density have their origin in the accumulated heat delivery (figure
\ref{fig:Heat}b) produced by the electrical current which have
`burned' into the material. This accumulated heating over several
time steps leads to a higher sintering activity of the hot regions and
therefore to a higher local density.

As a result, the simulation has proven to be in qualitative
  agreement with experiment, as far as 
the developing microstructure is concerned: Accumulated heat
fluctuations are
correctly given by the model and found as density fluctuations in
the experiment.

\begin{figure}[h]
\centering
  \includegraphics[width=0.95\textwidth]{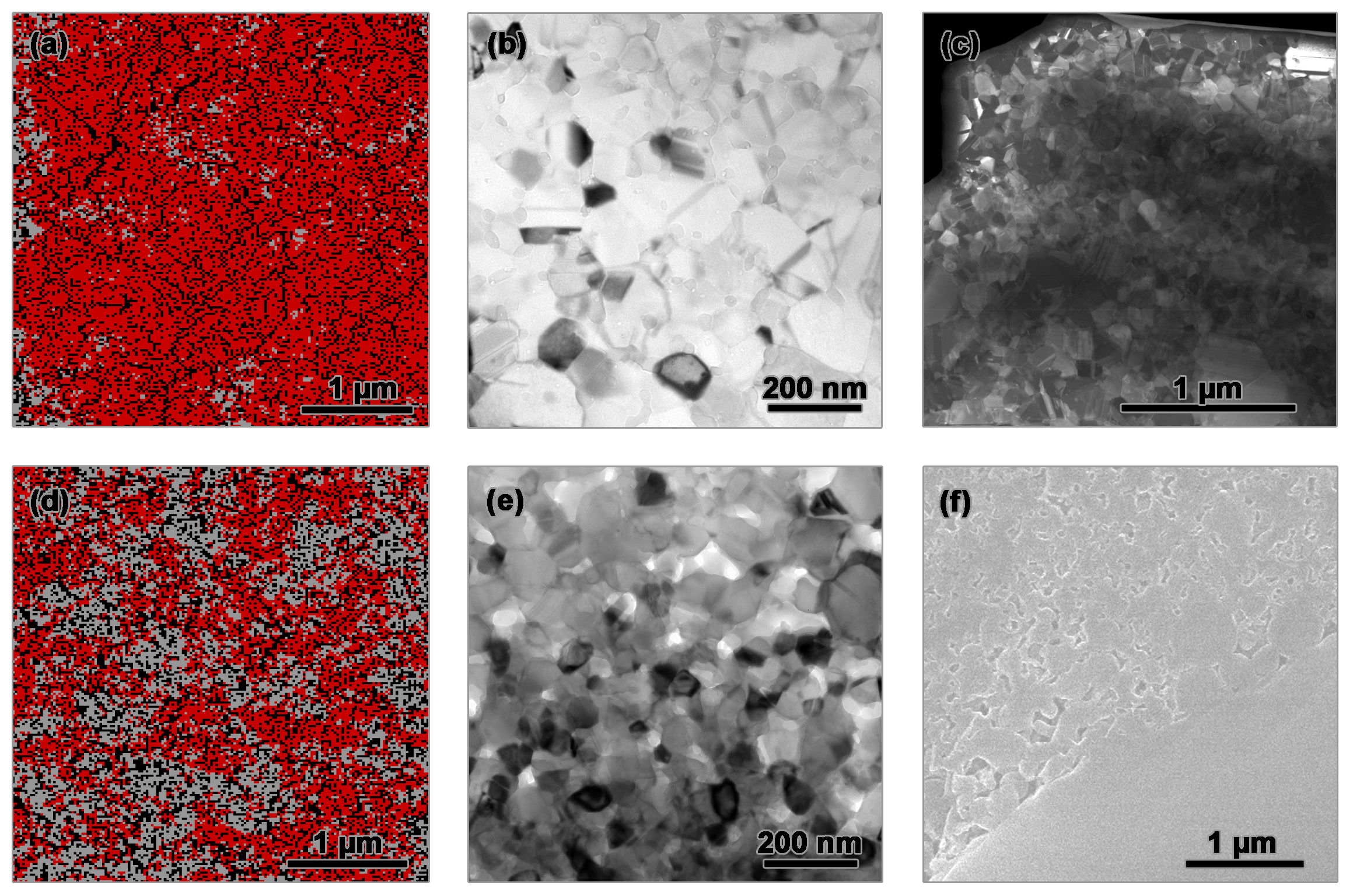}
  \caption{
    (a) Powder compacted with $m_{\tm{thres.}}=12$ which corresponds to a high sintering temperature.
    The red region consists of molten particles, while the grey dots are particles that are still in their original condition.
    (b) Scanning TEM (STEM) image of a nanosilicon with sintering
    temperature of \SI{1060}{\celsius}, density 97.2 \%. In STEM mode it can be seen, that the
    nanosilicon sample still has a high porosity.
    (c) Besides, SEM investigation shows that the samples has a good overall homogeneity --- inclusions seen in this image appear to be areas enriched with oxygen.
    (d) Powder compacted with $m_{\tm{thres.}}=24$ which corresponds to  a low sintering temperature.
    The color-coding is the same as in picture (a). The sample has a much higher amount of not-sintered powder (grey).
    Molten regions and not-sintered powder seem to cluster.
    (e) STEM image of a nanosilicon with sintering temperature of \SI{860}{\celsius}
    (density 95.0 \%), demonstrating porosity.
    (f) A SEM investigation showed that the sample is very inhomogeneous on the micrometer scale. Large, well sintered regions have
    formed and are found adjacent to regions of not-sintered powder.
  }
  \label{fig:Microstructure}
\end{figure}

In the following, the parameters of the simulation
are varied and compared with experimental investigations.
Therefore, the final microstructure of the simulation and the experiment are
compared in figure \ref{fig:Microstructure}.
Two threshold-values have been chosen, $m_{\tm{thres.}} = 12$ and
$m_{\tm{thres.}} = 24$, to represent a medium sintering temperature
(close to optimal compaction) and a low one,
respectively. The medium sintering temperature resulted in a
relatively homogeneous sample, both in simulation (figure
\ref{fig:Microstructure}a) and experiment (figure
\ref{fig:Microstructure}b and c). Figure \ref{fig:Microstructure}c
shows a nanosilicon which was sintered at \SI{1050}{\celsius} in
transmission. The nanocrystalline structure is preserved despite of
the density fluctuations due to the current paths as discussed
above.

For the low sintering temperature, a different microstructure is
expected according to the simulation. Regions which have been molten
and regions which were not heated sufficiently are found adjacent to
each other and seem to cluster as seen in figure
\ref{fig:Microstructure}d. This is found by a SEM analysis of the
accordingly sintered nanosilicon. On a scale of micrometer, very
well-sintered as well as bad-sintered regions have formed (figure
\ref{fig:Microstructure}f).

The reason for this may be attributed to the temperature gradients
created within the sample. If we assume that the viscous flow of the
particle physically always starts at the same threshold temperature
(e.g.\ the melting temperature of the nanoparticles), a low
sintering temperature (or high $m_{\tm{thres.}}$) means that the
temperature variation within the sample is higher because the mean
temperature is lower. This will necessarily create inhomogeneities.
Note, that these inhomogeneities cannot be reduced by longer
sintering times, nor will the density be increased considerably. The well
conducting paths have burned into the sample after a short time.
The bad-sintered regions will not carry the current to the
same amount as the well-sintered regions. Thus, longer sintering
times may even result in more extreme inhomogeneities. It can be
concluded: If the mechanism of densification is viscous flow,
homogeneous samples can only be obtained by choosing the sintering
temperature close to the threshold temperature of the viscous
flowing.

\begin{figure}[h]
\centering
  \includegraphics[width=0.5\textwidth]{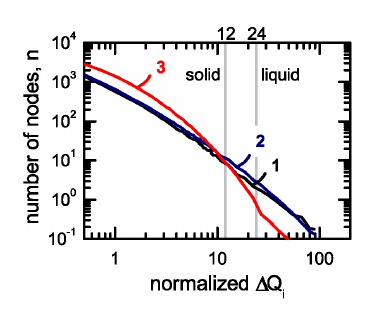}
  \caption{Number of nodes with a normalized heat $\Delta$Q$_i$.
  Curve 1 represents the initial distribution. The initial distribution shows
  no
  dependency of $m_{\tm{thres.}}$ and is only displayed for $m_{\tm{thres.}} = 12$
  . Curve 2 shows the heat distribution after 400 pulses
  for $m_{\tm{thres.}} = 12$, curve 3 the same situation for
  $m_{\tm{thres.}} = 24$. While there is not much change in the heat
  distribution of the sample in the case of a low threshold
  ($m_{\tm{thres.}} = 12$), it is found that the overall heat
  distribution
  changes drastically for higher thresholds ($m_{\tm{thres.}} = 24$). Only
  few particles are mobile (in their liquid phase), while a much
  higher number of nodes have a relatively low $\Delta$Q$_i$,
  creating an inhomogeneous temperature field within the sample.
  }
  \label{fig:Heatdis}
\end{figure}

Figure \ref{fig:Heatdis} shows a detailed analysis of the
temperature distribution (number of nodes $n$ at heat $\Delta Q_i$)
during the simulated sintering for $m_{\tm{thres.}} = 12$ and
$m_{\tm{thres.}} = 24$. It can be seen that the temperature
distribution for $m_{\tm{thres.}} = 12$ remains very close to the
initial configuration. Opposing, the distribution changes during
the process for $m_{\tm{thres.}} = 24$. More nodes with lower
$\Delta Q_i$ are found and less particles become mobile during each
time step. What does this mean for field assisted sintering?
There are only few hot spots created during the sintering if the
mean temperature is considerably lower than the threshold for
flowing ($m_{\tm{thres.}} = 24$). At these hot spots, the
microstructure will be reorganized only very locally, leading to
well-sintered areas which in turn will carry the current
afterwards.

For a homogeneously sintered sample, it is important
that current paths not only form, but disintegrate again. If the
mean temperature during the experiment is close to the threshold for
flowing ($m_{\tm{thres.}} = 12$), many particles become mobile in
each time step. Initially conducting current paths will thus be
destroyed by the
movement of the flowing particles. That is good for the formation of
a homogeneous microstructure. Of course, if a nanocrystalline
material is intended, the temperature profile has to be well
balanced: Too many flowing particles will lead to an unwanted
coarsening.

\section{Conclusion}

Concluding, we  used a network model to simulate the field assisted
sintering of highly doped silicon nanopowder and
compared the simulated results with experimental data. The evolution
of density during the sintering process qualitatively matched.
We found a good agreement between experiment and simulation with
respect to the local temperature distribution which is displayed as
density fluctuation in the microstructure of densely sintered
nanocrystalline silicon. So we captured some of the main aspects of
the FAST process with our simulation approach. As a consequence,
some information can be extracted as a guide for sintering dense,
homogeneous bulk applying field assisted liquid phase sintering:
After a very fast sintering phase in which the densification is
observed, the density saturates. Longer hold times, especially at
temperatures considerably below the threshold for flowing, will not
lead to a further densification but rather increase the
microstructural inhomogeneities within the sample. For the
developing microstructure, short sintering times at elevated
temperatures --- i.e.\ a temperature close to the threshold for
flowing --- result in best homogeneity combined with high density.

\section{Acknowledgments}
Financial support by the German Research Foundation (DFG) within the
Priority Program on nanoscaled thermoelectric materials, SPP 1386,
and within the Collaborative Research Center on nanoparticles from
the gas phase SFB 445, is gratefully acknowledged. Financial support
by the European Union and the Ministry of Economic Affairs and
Energy of the State of North Rhine-Westphalia in Germany in the
frame of an Objective 2 Programme (European Regional
Development,ERDF) and a Young Investigator Grant is gratefully
acknowledged. We thank M. Farle, department of experimental physics,
for the possibility to use the microscopy facilities.

\newpage

\bibliographystyle{unsrt}
\bibliography{Schwesig_paper_preprint}

\newpage
\section*{Figures}

\end{document}